# Reconfigurable inverse-designed phase-change photonics

Changming Wu[1], Ziyu Jiao[1], Haoqin Deng[1], Yi-Siou Huang[2,3], Heshan Yu[2,3], Ichiro Takeuchi[3], Carlos A. Ríos Ocampo[2,3] and Mo Li[1,4,*]

[1]Department of Electrical and Computer Engineering, University of Washington, Seattle, WA 98195, USA

[2]Department of Materials Science and Engineering, University of Maryland, College Park, MD 20742, USA

[3]Institute for Research in Electronics and Applied Physics, University of Maryland, College Park, MD 20742, USA

[4]Department of Physics, University of Washington, Seattle, WA 98195, USA

**Abstract**

**Chalcogenide phase-change materials (PCMs) offer a promising approach to programmable photonics thanks to their nonvolatile, reversible phase transitions and high refractive index contrast. However, conventional designs are limited by global phase control over entire PCM thin films between fully amorphous and fully crystalline states, which restricts device functionality and confines design flexibility and programmability. In this work, we present a novel approach that leverages pixel-level control of PCM in inverse-designed photonic devices, enabling highly reconfigurable, multi-functional operations. We integrate low-loss $Sb_2Se_3$ onto a multi-mode interferometer (MMI) and achieve precise, localized phase manipulation through direct laser writing. This technique allows for flexible programming of the photonic device by adjusting the PCM phase pattern rather than relying on global phase states, thereby enhancing device adaptability. As a proof of concept, we programmed the device as a wavelength-division multiplexer and subsequently reconfigured it into a mode-division multiplexer. Our results underscore the potential of combining inverse design with pixel-wise tuning for next-generation programmable phase-change photonic systems.**

[*] Corresponding author: moli96@uw.edu

**Introduction**

There are growing interests in programmable photonic integrated circuits (PICs) to enable highly flexible photonic networks for emerging applications such as optical computing[1,2], communication[3,4], optical interconnects for neural network accelerators[5–7], and quantum computing[8–10]. A critical factor in achieving these programmable PICs is the development of compact, high-performance programmable photonic components. Chalcogenide phase-change materials (PCMs) present a promising solution to realize programmable photonics due to their nonvolatile, reversible microstructural phase transition, ability for multilevel operation, and significant refractive index contrast between crystalline and armorphous phases ($\Delta n > 0.5$) [11–14]. PCMs, including $Ge_2Sb_2Te_5$, GSST, $Sb_2Se_3$, or $Sb_2S_3$, have been incorporated in Mach–Zehnder interferometers (MZIs)[15,16] and directional couplers[17,18] for phase modulation and in ring resonators for resonance tuning[19,20]. Typically, the functionality of these devices is determined by the phase state of the entire PCM film, whether in binary or multilevel configurations. In conventional designs, a thin film of PCM is deposited onto a photonic structure, with its phase controlled experimentally via thermal stimuli—either electrically using ocal heaters or optically through global tuning pulses. However, these programming methods offer limited spatial resolution, enabling phase control only over the whole PCM area. This constraint introduces challenges when designing programmable PCM photonic devices, as its programmable functions must correspond to fully amorphous and fully crystalline phases across the entire PCM layer. This constraint of device functionality to the phase condition of the entire PCM thin film restricts both its programmability and the full potential for reconfigurability, which is a missed opportunity given the significant index contrast PCMs offer. On the other hand, subdevice pixel-wise manipulation of PCM using laser writing has been demonstrated, allowing for localized programming[21]. However, the iterative process of switching individual PCM pixels is complex and inefficient, making it unsuitable for scalable implementations. To overcome the limitation, advanced design optimization techniques, including genetic algorithms[22,23], semi-analytical approaches[24,25], and notably, inverse design[26,27], can be applied to enhance the programmability of PCM-integrated photonic devices. Inverse design, in particular, has expanded the exploration of design spaces, allowing for the creation of devices with compact footprints and superior performance metrics[28]. Despite some progress, current implementations of inverse design for PCM devices primarily rely on patterning PCM materials into fixed patterns[29,30], while phase tuning is still achieved by switching the entire PCM region. The potential for subdevice pixel-wise control in inverse-designed PCM devices has yet to be fully leveraged, leaving substantial room for future advancements.

In this study, we present a reprogrammable, multi-functional photonic device by integrating the low-loss phase-change material $Sb_2Se_3$ onto a multi-mode interferometer (MMI). Unlike previously reported PCM devices, the function of our MMI is fully encoded by the inverse-designed phase pattern of the PCM. By employing direct laser writing in conjunction with inverse design optimization, we achieve pixel-level resolution in transferring the inverse-designed phase pattern to the $Sb_2Se_3$ thin film while preserving the integrity of other components. Remarkably, the functionality of this phase-change MMI device is fully rewritable, allowing for the erasure and reconfiguration of the function of the photonic device by reimprinting new phase patterns. As a demonstration, we successfully programmed the device to function as a wavelength-division multiplexing (WDM) device and subsequently reconfigured it into a mode-division multiplexing (MDM) device. Our results showcase the combined potential of inverse design and pixel-wise programming in phase-change photonic devices, paving the way for enhanced functionality and adaptability across different application scenarios.

**Results**

Figure 1a illustrates the schematic of the proposed reconfigurable multi-functional 1×2 MMI. Each MMI, occupying a footprint of 40×8.5 μm$^2$, is fabricated on a 330 nm-thick $Si_3N_4$-on-insulator substrate. A 30 nm-thick low-loss phase-change material, $Sb_2Se_3$, is sputtered onto the MMI and protected by a 200 nm-thick $SiO_2$ capping layer. The significant refractive index contrast between the two phases of $Sb_2Se_3$ (amorphous: $n_a$ = 3.285 and crystalline: $n_c$ = 4.050, respectively) allows efficient evanescent coupling of light within the MMI with the $Sb_2Se_3$ thin film. As a result, the light undergoes a phase modulation that depends on the local structure of $Sb_2Se_3$. By inducing localized perturbations, the phase pattern of $Sb_2Se_3$ on MMI shapes the wavefront during light propagation, altering the amplitude and phase distribution of the propagating light. This capability allows the MMI to execute diverse functions with minimum loss penalty. In particular, we can associate the specific function of the MMI device with a distinct binary $Sb_2Se_3$ phase pattern within the "inverse-design region" highlighted by the red dashed box in Figure 1b. Furthermore, as depicted in Figure 1c, the MMI's function is entirely encoded within the PCM thin film, while the other components remain unchanged. The reversible phase change of $Sb_2Se_3$ allows for convenient erasure and recreation of the phase pattern, thereby lifting the limitation of binding device functionality to the phase condition of the entire PCM thin film. This enables a more flexible approach to designing the functionality of PCM photonic devices.

Programming the MMIs involves designing and patterning processes. First, we determine the MMI's function and design the corresponding $Sb_2Se_3$ phase pattern. To achieve this, we employ a topology optimization approach to inverse-design the phase pattern of the $Sb_2Se_3$ thin film [31,32]. In our simulations, we segment the $Sb_2Se_3$ thin film within the "inverse-design region" into an array of pixels, each measuring 100 nm × 100 nm × 30 nm, while maintaining the other components of the MMI unchanged during optimization. To avoid unachievably fine features, we also limit the minimum feature size in the designed patterns to be larger than 500 nm. The optimization process initiates with a linear parametrization, allowing the dielectric permittivity $\varepsilon_i$ of each pixel to continuously vary in the range of ($n_a$, $n_c$). Subsequently, the phase pattern is updated using the steepest descent method. Through this gradual optimization process, the permittivity of the pixels transitioned from an intermediate value in ($n_a$, $n_c$) toward a binary value of $n_a$ or $n_c$. Consequently, the phase map evolved from a grayscale phase pattern to the binary phase pattern of $Sb_2Se_3$, as illustrated in Figure 2a.

Second, the corresponding phase pattern is directly laser written onto the MMIs using a commercial 405 nm laser direct lithography system (Heidelberg DWL 66+, 405 nm laser). The exceptional speed and accuracy of the DWL system enable efficient and effective patterning of the target MMI device. During the experiment, we began with the $Sb_2Se_3$ thin film in the completely $cSb_2Se_3$ phase and wrote phase patterns directly onto the MMI by quenching $cSb_2Se_3$ into $aSb_2Se_3$ phase using optical heating[33,34]. Figure 2b displays a series of rectangular $aSb_2Se_3$ patterns with widths ranging from 200 nm to 2 μm written on a blank $cSb_2Se_3$ film, demonstrating a minimum feature size of ~300 nm achieved through laser writing. We note that this resolution reaches the subwavelength regime and is much smaller than the resolution achieved in previous works[33,35,36].

Furthermore, the pattern can be readily erased through an annealing process, either by local heating using a continuous-wave laser or global heating to 180 °C using a hotplate (see Figure 2c). By selectively erasing and rewriting the patterns on MMIs, the function of the circuit is reprogrammed. This technique gives us a toolkit for writing high-finesse functional patterns into the phase-change films and modifying or erasing them as required.

Using the direct writing technique, we first demonstrate programming the MMI as a mode demultiplexer (MDM). The $Sb_2Se_3$ phase pattern (Figure 2e), generated with the inverse design approach, was precisely aligned (misalignment < 125 nm) and written onto the MMI. As shown in Figure 2d, the laser-written phase pattern closely resembles the design with minimal error. Subsequently, we characterized the MDM performance of the MMI device. Figure 3a shows an optical microscope image of the complete photonic circuit and a sketch of the measurement setup. A multimode waveguide, supporting both $TE_0$ and $TE_1$ modes, is connected to the input of the MMI, while both output ports (port 2 and port 3) are linked to single-mode waveguides for measurement. The incident light is coupled to the device via either port 1 or port 4 and passes through a mode combiner constructed with an asymmetric directional coupler. This combiner selectively convertthe incoming $TE_0$ mode from incident port 4 to the $TE_1$ mode while maintaining the $TE_0$ mode from input port 1. The design and characterization of the mdoe combiner is included in Supplementary Information. As a result, the input mode to the MMI is determined based on which port is used (port 1 for $TE_0$ or port 4 for $TE_1$), which is selected using a 1×2 optical switch. Finally, the output power from both output ports (port 2 and port 3) was measured to determine the respective transmission coefficients of each mode.

Figures 3b and 3c illustrate the MDM's functionality, routing the $TE_0$ mode to port 2 and the $TE_1$ mode to port 3 over a wavelength range of 1500 to 1600 nm. Figures 3d and 3e compare the simulated and measured transmission spectra of the two modes. In the simulation, this MDM device demonstrates a modal extinction ratio surpassing 15 dB for both mode channels over a bandwidth wider than 100 nm. The experimental results exhibit a extinction ratio of > 10 dB for both mode channels, lower than the simulated values. We attribute the underperformance to fabrication imperfections, misalignment, and roughness in the phase pattern. Additionally, in our measurements, the grating couplers constrain the insertion loss and operational bandwidth. Furthermore, we fabricated three identically devices, which demonstrated similiar performance

(see Supplementary Information), manifesting the consistency of the design and fabrication approach.

We then demonstrate how this technique enables a more flexible approach to designing the functionality of PCM photonic devices, moving beyond the constraint of binding device functionality to the phase condition of the entire PCM thin film. We reprogram the functionality of the MMI from an MDM to a wavelength demultiplexer (WDM). To do so, we erased the previous $Sb_2Se_3$ phase pattern and rewrote a newly designed phase pattern onto the same MMI, as shown in Figures 4a to 4c. To accomplish the WDM function, we redesigned the phase pattern within the "inverse-design region" using the same method described earlier (Figure 4d). Figures 4e and 4f show the simulated electric field distribution at two operating wavelengths of 1475 nm and 1625 nm, respectively. The WDM separates the $TE_0$ mode of the input waveguide into the two output ports according to their wavelength: 1475 nm at port 2 and 1625 nm at port 3. Figures 4g and 4h plot the measured and simulated transmission spectra for the WDM, respectively. The simulation shows a high extinction ratio of > 15 dB at the center wavelengths. The experimental results show a lower extinction ratio of >6 dB measured at wavelengths of 1465 nm and 1695 nm, which is limited by the range of the tunable laser source used. The reduced performance can again be attributed to fabrication imperfections and pattern misalignment. Nevertheless, the results successfully demonstrate a complete change of the device's functionality, from an MDM to a WDM, by recreating the phase pattern of the $Sb_2Se_3$ thin film.

**Conclusion**

In summary, our work demonstrates the versatility of a multi-functional photonic device achieved through the integration of low-loss phase-change material $Sb_2Se_3$ onto a silicon nitride photonic integrated circuit, where functionality is encoded in the phase pattern of the $Sb_2Se_3$ thin film unlike previous design logic for phase-change photonics that binding device functionality to the phase condition of the entire PCM thin film. Leveraging direct laser writing and inverse design optimization, this novel phase-change MMI device offers excellent flexibility for a single photonic device. We successfully programmed this device to operate as multiplexers for either wavelengths or spatial modes. Utilizing photonic inverse design ensured high extinction ratios and broadband response in both configurations. Moreover, there is potential for further exploration of the application scenario by adopting a multi-level grayscale design instead of the current binary design and cascading multiple stages of such photonic structures to enable a diverse range of complex reconfigurable photonic systems.

**Data Availability Statements**

The data that support the findings of this study are available from the corresponding author upon reasonable request.

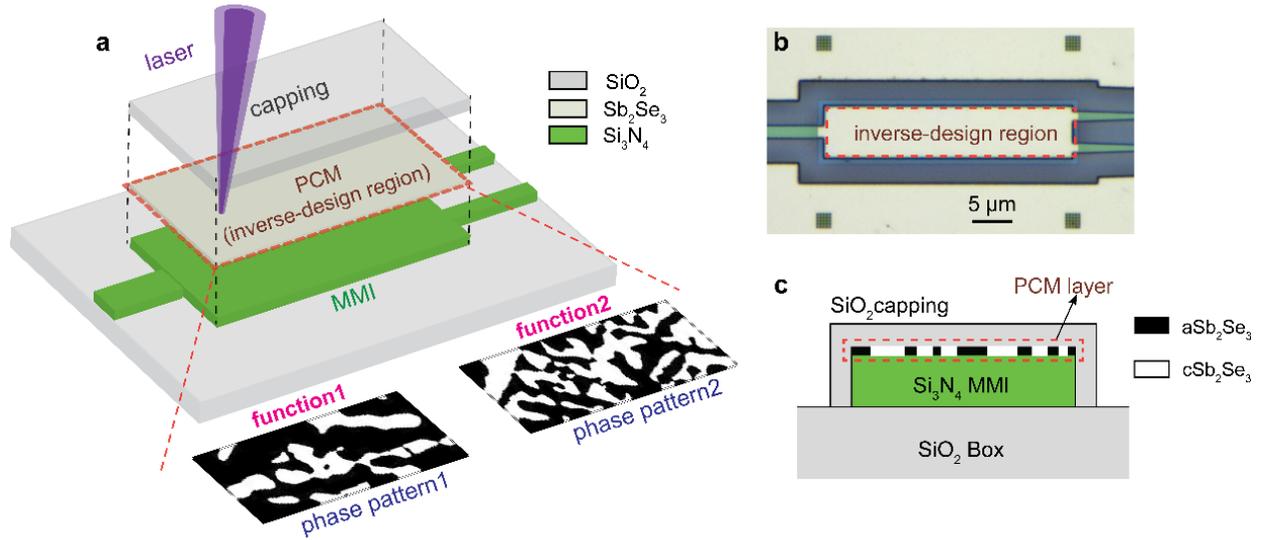

**Figure 1 Schematic of the laser writing of multi-functional photonic MMI. a.** Illustration of programming a Sb$_2$Se$_3$ thin film integrated MMI using a focused laser beam. The commercial direct laser writing system (Heidelberg DWL66+) is used to write/erase various phase patterns onto the Sb$_2$Se$_3$ thin film, which determines the function of the MMI. **b.** The top-view optical image of an as-fabricated 40×8 μm$^2$ MMI. The device's functionality is defined by the Sb$_2$Se$_3$ phase pattern in the "inverse design region" (red dashed box). **c.** The cross-sectional schematic of the MMI device with patterned Sb$_2$Se$_3$. To change the function of the device, the phase pattern in the Sb$_2$Se$_3$ thin film is recreated, while all the other components remain unchanged.

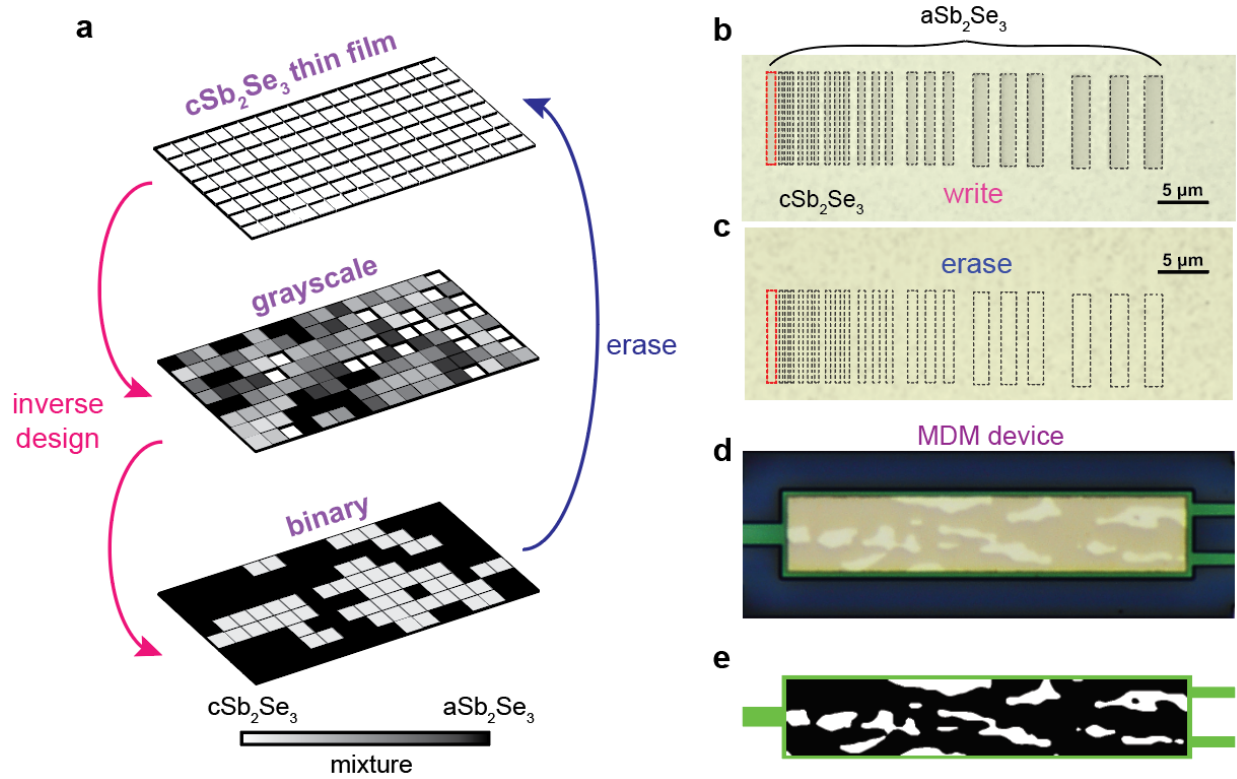

**Figure 2 Inverse design of the Sb₂Se₃ phase pattern. a.** Schematic of the inverse design approach to generate phase pattern for the Sb₂Se₃ MMI. The optimization is initiated with a fully crystalline Sb₂Se₃ thin film, and then the amorphous-crystalline spatial distribution is inverse-designed using the topology optimization method, resulting in a binary phase pattern. White and black areas indicate the crystalline and amorphous Sb₂Se₃, respectively. **b.** Optical image of aSb₂Se₃ rectangular array written on cSb₂Se₃ thin film for testing the writing resolution. The minimum feature size achieved is 300 nm. **c**. The test pattern in (**b**) is erased back to the cSb₂Se₃ phase by thermal annealing. The area marked by dashed lines indicates the initial patterns. **d.** Optical images of an inverse-designed pattern for the mode demultiplexer written in Sb₂Se₃. **e.** The corresponding inverse designed, binarized Sb₂Se₃ phase pattern for the mode demultiplexer.

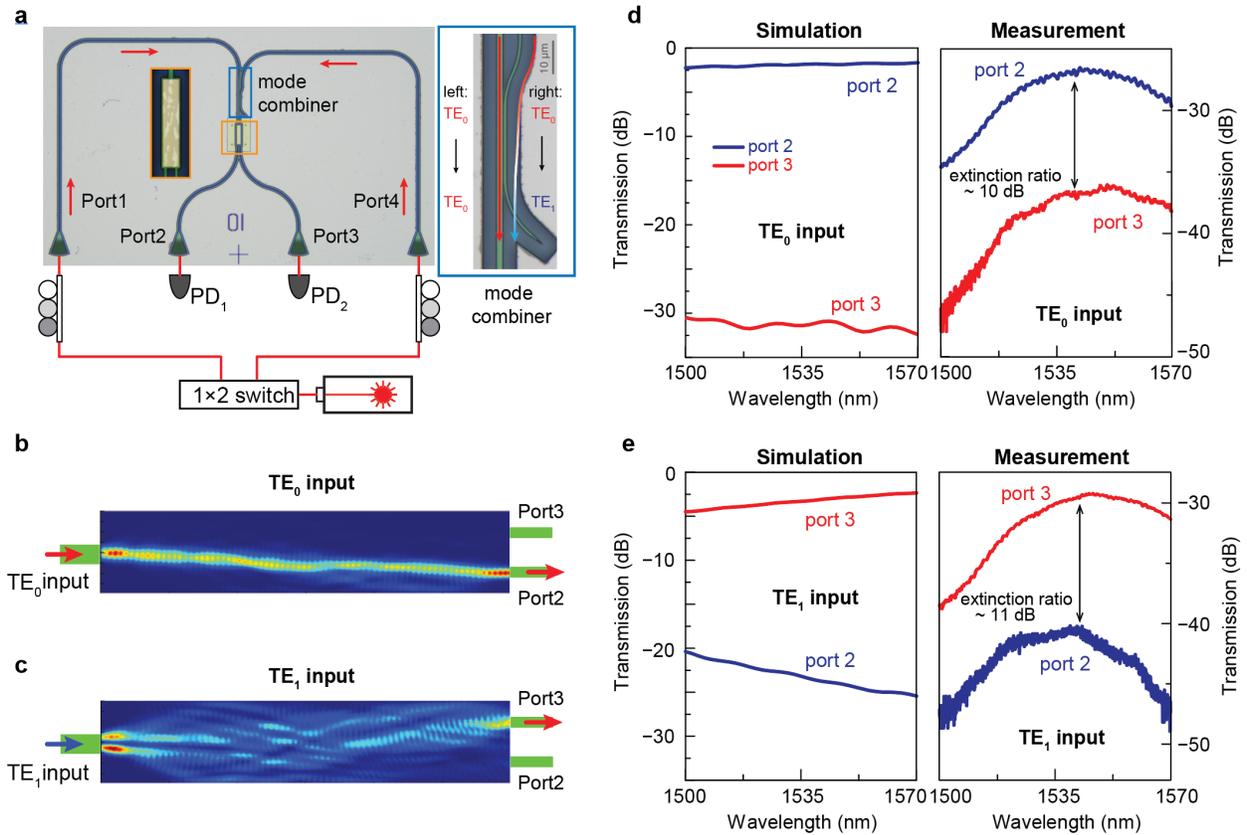

**Figure 3. MMI programmed as a mode demultiplexer. a.** The optical setup for the MDM device measurement. An optical switch and a mode combiner control the input mode, $TE_0$ or $TE_1$, to the MDM. Inset: the optical image of the mode combiner. **b.** and **c.** The simulated $|E|^2$ distribution in the MDM with $TE_0$ mode input (**c**) and $TE_1$ mode input (**c**) at the wavelength of 1550 nm. **d** and **e**. The simulated (left column) and measured (right column) transmission spectrum of the mode demultiplexer when the $TE_0$ mode (**d**) or $TE_1$ mode (**e**) is input. The lower extinction ratio is attributed to fabrication imperfections and laser writing misalignment.

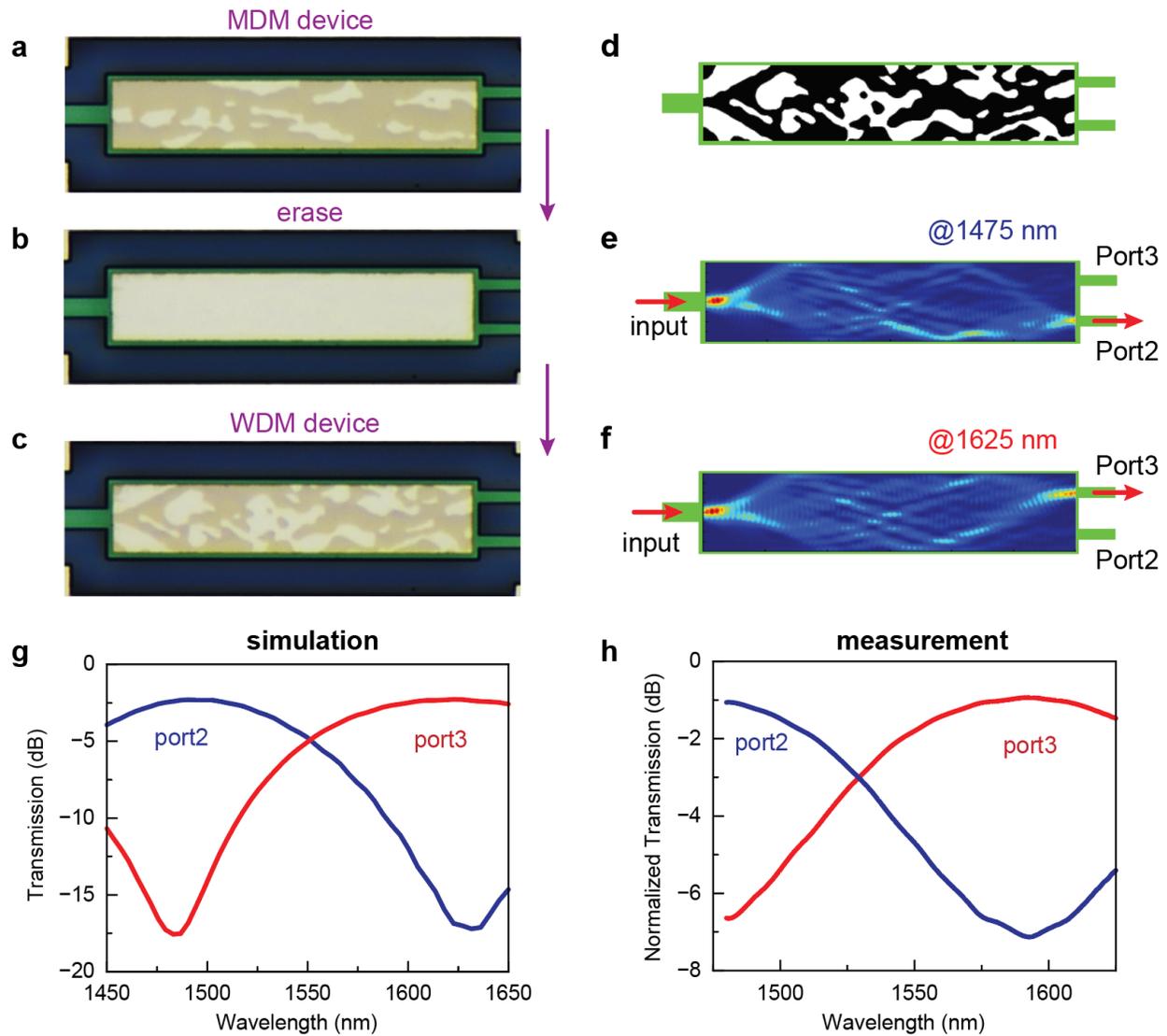

**Figure 4 Reprogramming to a wavelength demultiplexer. a to c.** Optical images showing the steps that the MDM device (**a**) is first erased to the full crystalline cSb$_2$Se$_3$ thin film(**b**) and then reprogrammed as a WDM (**c**). **d.** Inverse-design binarized Sb$_2$Se$_3$ phase pattern for the WDM. White and black areas indicate the crystalline and amorphous Sb$_2$Se$_3$, respectively. **e** and **f.** Simulated |E|$^2$ distribution of the WDM device operating at 1475 nm (**e**) and 1625 nm (**f**). **g** and **h.** Transmission spectra to output ports 2 and 3 obtained from simulation (**g**) and measurement (**h**).